\documentclass[showpacs,preprintnumbers,superscriptaddress,amsmath,reprint,aps,pra,twocolumn,longbibliography]{revtex4-1}

\usepackage{bm}
\usepackage{graphicx} 
\usepackage{color}
\usepackage{amsmath}
\usepackage{booktabs}
\usepackage{epstopdf}
\usepackage{float}
\usepackage{color}

\newcommand{\EEA}{\end{eqnarray}}
\newcommand{\EEAN}{\end{eqnarray*}}

\newcommand{\cev}[1]{\reflectbox{\ensuremath{\vec{\reflectbox{\ensuremath{#1}}}}}}

\begin{document}
 
\title{Anisotropic blockade using pendular long-range Rydberg molecules }

\author{Matthew T. Eiles}

\affiliation{Department of Physics and Astronomy, 
Purdue University,  47907 West Lafayette, IN, USA}

\affiliation{Kavli Institute of Theoretical Physics, University of California, Santa Barbara, Santa Barbara, California 93106, USA}

\author{Hyunwoo Lee}

\affiliation{Department of Physics and Astronomy, 
Purdue University,  47907 West Lafayette, IN, USA}

\author{Jes\'{u}s P\'{e}rez-R\'{i}os}

\affiliation{Department of Physics and Astronomy, 
Purdue University,  47907 West Lafayette, IN, USA}

\author{Chris H. Greene}

\affiliation{Department of Physics and Astronomy, 
Purdue University, 47907 West Lafayette, IN, USA}

\affiliation{Kavli Institute of Theoretical Physics, University of California, Santa Barbara, Santa Barbara, California 93106, USA}

\affiliation{ Purdue Quantum Center, Purdue University, West Lafayette, Indiana 47907, USA.}

\date{\today}

\begin{abstract}
We propose an experiment to demonstrate a novel blockade mechanism caused by long-range anisotropic interactions in an ultracold dipolar gas composed of the recently observed ``butterfly'' Rydberg molecules.  At the blockade radius, the strong intermolecular interaction between two adjacent molecules shifts their molecular states out of resonance with the photoassociation laser, preventing their simultaneous excitation. When the molecules are prepared in a quasi-one-dimensional (Q1D) trap, the interaction's strength can be tuned via a weak external field. The molecular density thus depends strongly on the angle between the trap axis and the field. The available Rydberg and internal molecular states provide a wide range of tunability. 
\end{abstract}

\maketitle
\section{Introduction}
Ultracold dipolar gases provide an ideal environment for the study of novel ultracold chemical reactions or quantum chaos in non-linear dynamical systems~\cite{chem1,chem2,chem3,PRX,Nature}, the design of robust quantum information protocols~\cite{rmp,DeMille,Wie}, and investigations of universality in few- and many-body physics~\cite{fbody1,fbody2,ReviewZoller,MB2,MB3,MB4,MB5,MB6,MB7,MBextra,MBextra2,Pupillo}. These promising applications hinge on the premise that regimes exist where the dipole-dipole interaction (DDI) is the dominant force in the system. Recent experiments have partially achieved this in three different systems: ultracold polar molecules \cite{KRb,NaK,NaRb,RbCs}, ultracold lanthanide atoms with large magnetic moments~\cite{Nature,PRX,Ferrofluid,PfauDroplet,Lev,Cr}, and Rydberg atoms in external fields \cite{Noel, Ravets,Pupilloryd}. These are challenging experiments: large external fields ($\sim10^4$V/cm) are required to align polar dimers \cite{alkalidimers}, and their production in their rovibrational and hyperfine ground state is a titanic experimental effort \cite{KRb}.  On the other hand, the small atomic magnetic moments require the reduction of the atom-atom interaction using Feschbach resonances \cite{zollerpolar,pfautune}. The properties of polar dimers and magnetic atoms are rarely tunable.  Rydberg atoms suffer from comparatively short lifetimes, and only interact through the purely isotropic van der Waals interaction unless dipole moments are induced via an external field \cite{Marcassa} or a F\"{o}rster resonance \cite{SaffmanBlockade}. These techniques require detailed knowledge of the Stark-induced avoided crossings and delicate control of applied electric fields. 
 
\begin{figure}[b]
{\normalsize 
\includegraphics[scale =0.41]{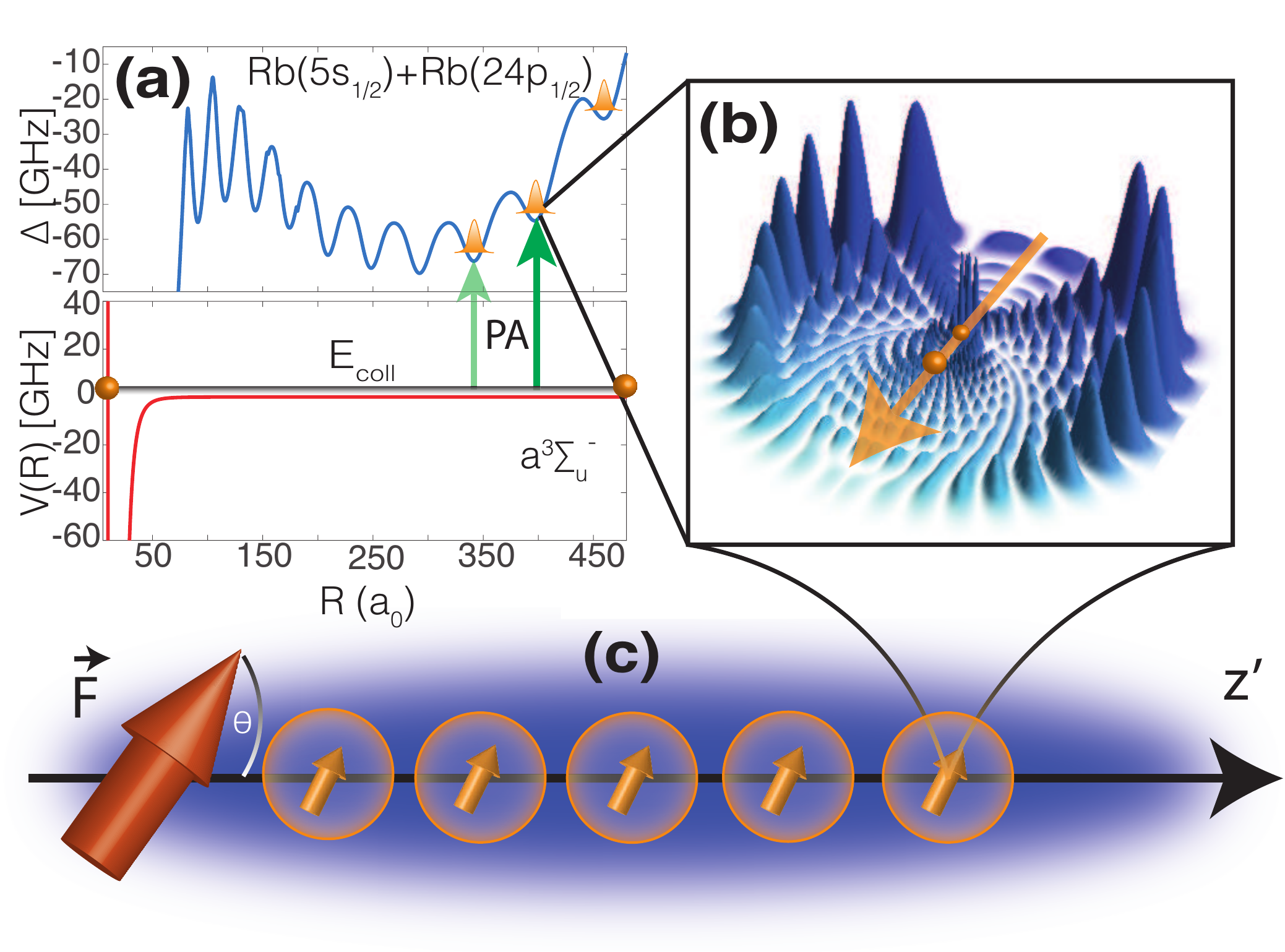}
}
\caption{a) The proposed single-photon photoassociation scheme. The target bound state is selected by the laser frequency. b) The electronic probability density $\rho|\Psi(\rho,z)|^2$ in cylindrical coordinates. The two atoms (sphere) and electric dipole (arrow) are shown. c) The experiment: aligned pendular butterfly molecules in a Q1D cloud. The blockade radius prevents simultaneous excitation of two close molecules.}
\label{fig:expsetup}
\end{figure}
The recent observation of ultra-long-range ``butterfly" Rydberg molecules  \cite{Butterfly} suggests a unique hybrid system combining the promising properties of these systems. The properties of these {\it homonuclear} molecules are readily tunable over a large range of values, and they possess gigantic permanent electric dipole moments (PEDMs). This is due to their unique bonding mechanism: the Rydberg electron repeatedly scatters off a nearby perturbing atom, creating a novel chemical bond between these two atoms. These molecules therefore possess the exaggerated characteristics of Rydberg systems and the internal structure and polar attributes of molecules, without many of the experimental challenges present in these other systems.   This article explores the dipolar physics of butterfly molecules in a Q1D array of molecules aligned by a weak ($<1$V/cm) external electric field applied at an angle $\theta$ relative to the longitudinal trap axis. Intermolecular forces prevent the resonant excitation of two molecules within the butterfly blockade radius $R_b(\theta)$, where the anisotropic intermolecular potential, $V(R_b,\theta)$, exceeds the laser bandwidth, $\Gamma$ ( $\sim0.5$ MHz). $R_b(\theta)$ can be tuned by the applied field: at the ``magic angle'' satisfying $P_2(\cos\theta_M)=0$ ($P_L$ is the Legendre polynomial of order $L$) the DDI vanishes and higher-order terms in the molecular interaction, namely the quadrupole-quadrupole/dipole-octupole and van der Waals, dominate. Thus,  the molecular density reveals these interactions via its dependence on $\theta$. The high tunability and exaggerated scales of this proposal has implications in studies of the polaron problem, angulon interactions, tunable interactions, and crystalline phases. 


  The organization of this article follows the hierarchy of energy scales in this system. First, in section II the properties of butterfly molecules, which are bound by several GHz and have electronic/vibrational spacings of a few GHz are discussed. Next, the pendular states are calculated within the rigid-rotor approximation; their energy splittings are a few tens of MHz.  Section III details the perturbative calculation of the intermolecular interactions; their maximum strength for our considerations is restricted by $\Gamma$ ($\sim 500$kHz). Once this interaction is calculated the density of molecules is readily obtained as a function of $\theta$, and these results are given in section IV. 
  \section{pendular states of long-range butterfly molecules}

Butterfly molecules consist of a Rydberg atom bound to a ground state atom \cite{Greene2000}. The electron-atom scattering process is described using contact pseudopotentials proportional to
  $-\tan\delta_L/k^{(2L+1)}$, where $\delta_L$ is the elastic phase shift for partial wave $L$ and $k$ is the electron's momentum \cite{Fermi,Omont}. The contributions from partial waves $L>0$ typically vanish at ultracold temperatures. However, the presence of a $P$-wave shape resonance causes $\tan\delta_1$ to diverge, resulting in a potential curve descending from the $n$ atomic line to below the $(n+2)p$ Rydberg level, which is red-shifted by its quantum defect $\mu_p \approx 2.65$ \cite{HamiltonGreeneSadeghpour}.  Level repulsion from the $n-1$ hydrogenic manifold constrains the divergence \cite{HamiltonGreeneSadeghpour,KhuskivadzePRA,KhuskivadzeJPB}.  The potential curve oscillates as a function of the internuclear distance $R'$ and molecules form in the potential wells mirroring the $p$-wave wave function \cite{Butterfly}; the relatively high $p$-character found in the butterfly wave function allows for photoassociation in a dense condensate through a single photon process.  Neglecting spin, the pseudopotentials are
\begin{equation}
V_{pp}(\vec R',\vec r)=2\pi a_S\delta(\vec r - \vec R') + 6\pi a_P\cev\nabla\cdot\delta(\vec r - \vec R')\vec\nabla. 
\label{fermieqn}
\end{equation}
\begin{figure}[b]
{\normalsize 
\hspace{-10pt}\includegraphics[scale =0.36]{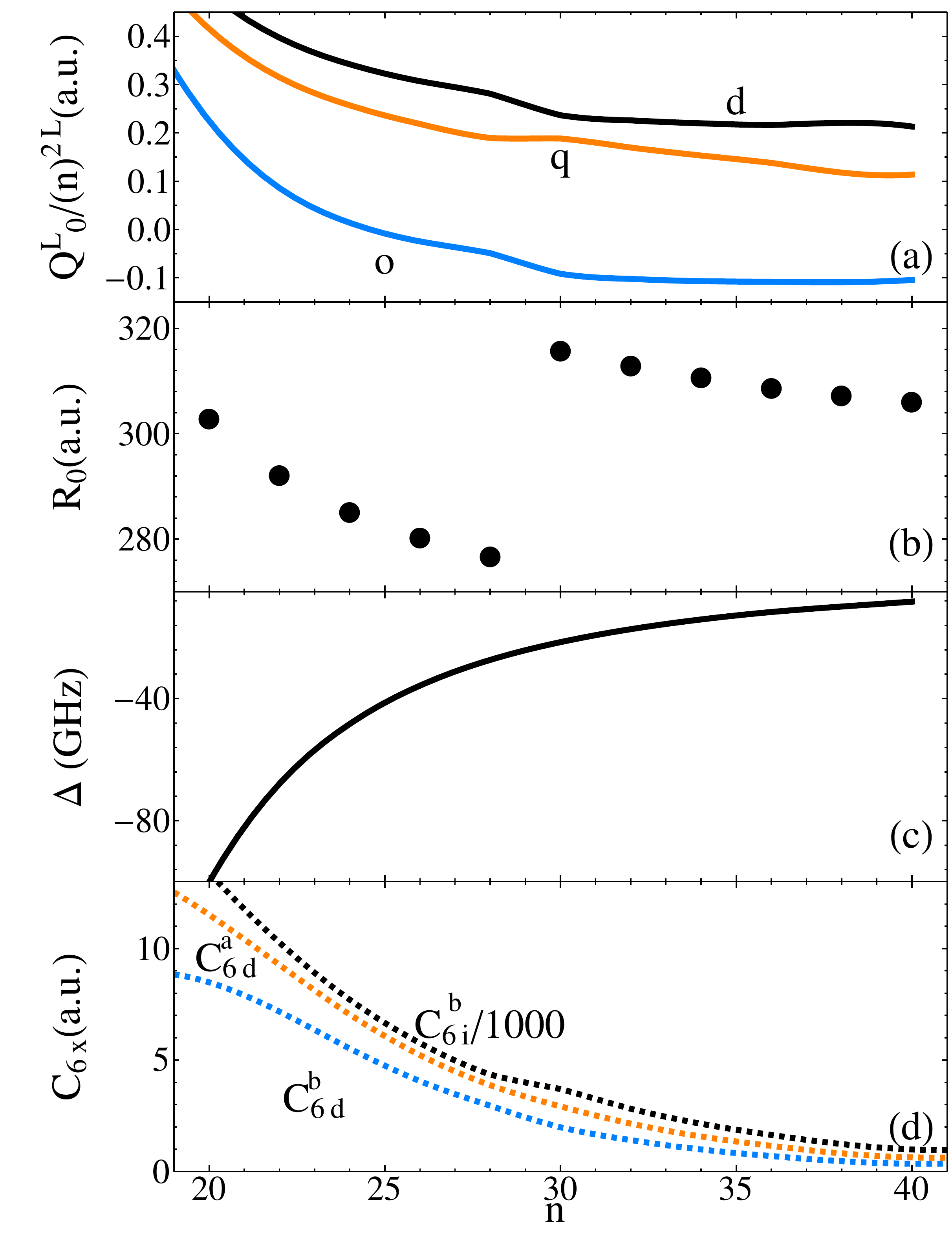}
}
\caption{ Properties of the deepest pendular state as a function of $n$: a)  multipole moments, labeled $d = Q^1_0/n^{2}$, $q = Q^2_0/n^4$, and $o = Q^3_0/n^6$;  b) bond lengths; c) binding energies; d) Relevant van der Waals coefficients at $F=1$V/cm. ( Eq. \ref{eqn:interactionresults}).}
\label{fig:ndependence}
\end{figure}
Eq. \ref{fermieqn} generalizes to include spin-dependent scattering parameters \cite{Matt}. The full Hamiltonian $H = H_0 +H_{SO} + V_{pp} + H_{HF}$ includes the Rydberg electron's Hamiltonian, $H_0$, the Rydberg spin-orbit interaction $H_{SO}$, and the hyperfine splitting of the ground state atom, $H_{HF}$  \cite{AndersonPRA,Matt,Markson}.  Diagonalization of $H$ reveals binding energies, bond lengths, and multipole moments \cite{Matt}. A typical potential curve and vibrational states are plotted in Fig. \ref{fig:expsetup}a. The electronic state associated with this potential is shown in Fig. \ref{fig:expsetup}b and consists of a mixture of Rydberg states that maximizes the radial derivative of the wave function at $\vec r = \vec R'$.  Fig. \ref{fig:ndependence} displays various molecular parameters of the ground state as a function of $n$.  The reduced multipole moment of molecule $X$ averaged over the vibrational molecular wave function is $q(L_X)=Q_0^{L_X}/n^{2L_X}$.   Greater tunability is possible if different molecular states are excited; the dipole moment varies over a factor of $\sim 3$ for different vibrational states.

\begin{figure}[t]
{\normalsize 
\includegraphics[scale =0.15]{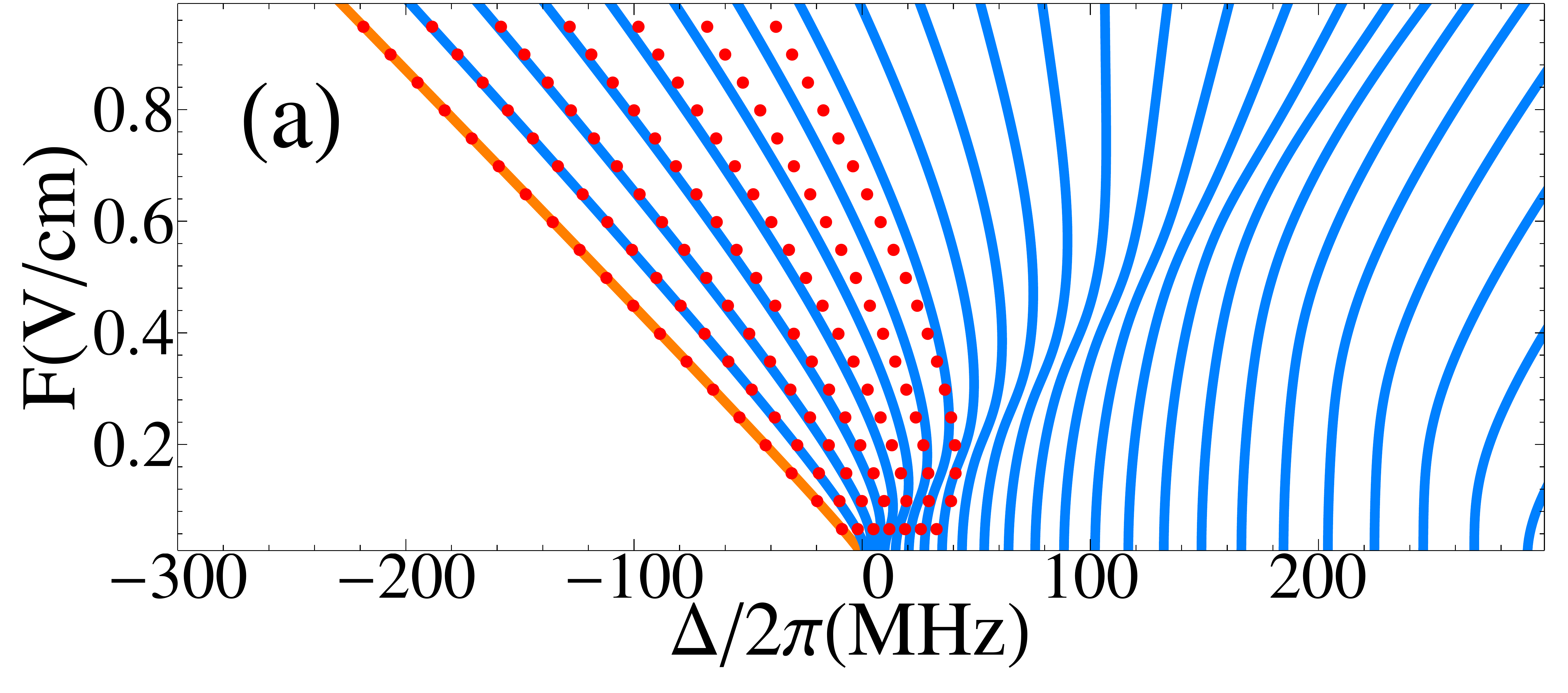}\\
\vspace{-8pt}
\includegraphics[scale =0.15]{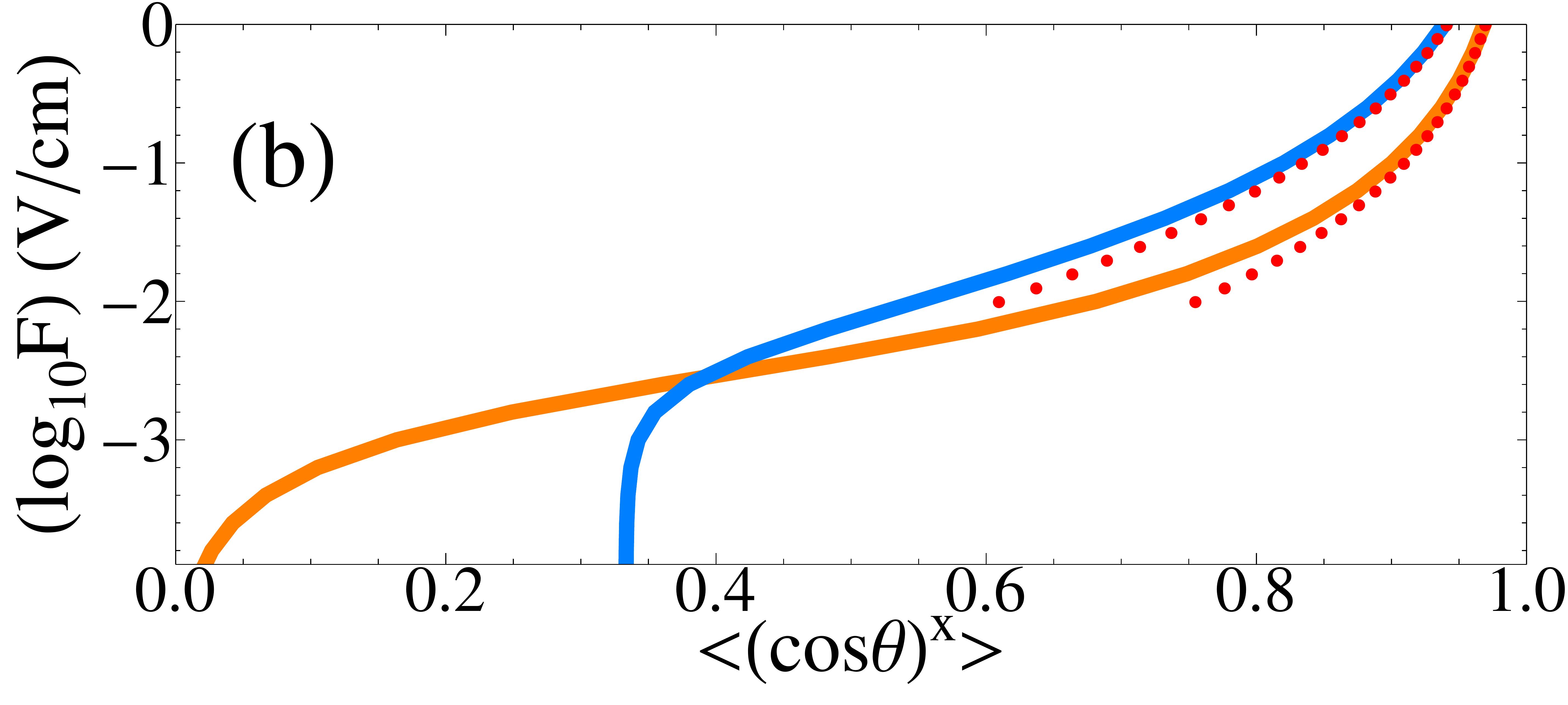}
\vspace{-20pt}
}
\caption{(a) Stark spectrum of the $n=24$ pendular states, showing the energy shift $\Delta$ as a function of the applied electric field for $M_{N}=0$. The red points correspond to the two-dimensional harmonic oscillator approximation. (b) Orientation $(x=1)$ (orange) and alignment $(x=2)$ (blue) of the lowest pendular state [orange line in (a)].
 }
\label{fig:pendular}
\end{figure}
In the absence of external fields, polar molecules  rotate freely with random orientations. Application of a field shifts the molecular energy through the dipole-field coupling $-\vec{d}\cdot \vec{F}$, where $\vec{F}$ is the electric field and $\vec{d}$ is the molecular PEDM. Setting the quantization axis parallel to the electric field, the molecular Hamiltonian is $
H_\text{mol}=B_e\hat{N}^2-dF\cos\theta
$,
with rotational constant $B_e$ and rotational angular momentum operator $\hat{N}$ ~\footnote{We work in Hund's case b}. When the dimensionless parameter $\omega=\frac{dF}{B_e}$ is large ($\omega\sim 10^2-10^3$ for field strengths $\sim 1$ V/cm, four orders of magnitude lower than needed for typical heteronuclear molecules ~\cite{KRb,Butterfly}), the rotational states become trapped in a nearly harmonic potential and are called pendular states in analogy with the harmonic oscillator ~\cite{Rost,Friedrich}. These states are found by diagonalizing $H_\text{mol}$ in the basis of spherical harmonics $Y_{NM_{N}}(\theta,\phi)$; in the large $\omega$ limit two-dimensional harmonic oscillator eigenfunctions are excellent approximations (red points in Fig. \ref{fig:pendular}a are calculated using this approximation; the appendix describes this approximation in further detail). The pendular states $|\tilde NM_N\rangle$ are characterized by their librational state, $\tilde N$, and the good quantum number $M_N$. The ground state, $|00\rangle$, is the most aligned pendular state. Fig. \ref{fig:pendular}a shows the resulting Stark spectrum for the $n=24$ case.  Fig. \ref{fig:pendular}b shows  $\langle \cos{\theta} \rangle$ and $\langle \cos^2{\theta} \rangle$.

\section{Calculation of the intermolecular interaction}

Fig. \ref{fig:expsetup}c depicts a sketch of the geometry of our proposal. The electric field points in the lab frame's $z$ axis; a strongly confining potential in the $x$ and $y$ dimension creates a Q1D chain of Rb atoms in the $z'$ direction \cite{Ryd1D}, where $\cos\theta = \hat z\cdot\hat z'$. The laser is tuned to excite the $|00\rangle$ pendular butterfly state, although greater parameter ranges can be explored via the internal structure associated with other rotational states.  Two molecules are separated by a distance $R$ that depends on the atomic density and the long-range intermolecular interaction.

The intermolecular interaction, given by the two-center multipolar expansion of the Coulomb force \cite{MargenauInt,Stone,Gronenboom,Zanchet,Flannery,vanderAvoird1980,SuppMat}, is valid for $R> 2(\sqrt{\langle r_A^2\rangle} + \sqrt{\langle r_B^2\rangle})$ \cite{LeRoy}, which is satisfied for all distances studied here. It is calculated to order $R^{-6}$, requiring second order perturbation theory. The zeroth order wave functions are the pendular states obtained by diagonalizing $H_\text{mol}$, whose energy spacings are typically an order of magnitude smaller than the vibrational or electronic spacings. We thus neglect contributions from other electronic or vibrational levels in the second-order sum over intermediate states, which should introduce errors in the $1/R^6$ potentials of 10\% or less due to the larger energy separations.   The overall $n$-scaling of the multipole moments factors out, along with an additional $n^3$ for the second-order terms from their energy denominators. This gives
\begin{align}
\label{eqn:interactionresults}
&V(R,\theta)=\\\nonumber & -\frac{2C_3d^2n^4}{R^3}P_2(x)-\frac{8n^8}{R^5}P_4(x)\left(C_{5a}do-C_{5b}q^2\right) \\&\nonumber-2\frac{4d^4n^{11}}{R^6}\left(C_{6i}^a[P_2(x)]^2+C_{6i}^b\frac{(xy)^2}{4}\right)\\
&-\frac{4d^4n^{11}}{R^6}\left(C_{6d}^a[P_2(x)]^2+\frac{C_{6d}^c}{4}y^4+C_{6d}^b\frac{(xy)^2}{4}\right),\nonumber
\end{align}
where $x=\cos\theta,y=\sin\theta$, and $C_q>0$.  We obtain $C_3\sim0.95$,  $C_{5a}\sim 0.83$, and $C_{5b}\sim 0.63$ for the first-order coefficients, independent of $n$ within 2\% over the range $n=20-40$. The second-order coefficients vary slowly with $n$ as seen in Fig. \ref{fig:ndependence}d. The van der Waals terms have been split into induction $(C_{6i})$ and dispersion $(C_{6d})$ terms \cite{vanderAvoird1980}. Several unexpected properties of these coefficients emerge: the independence of the first-order coefficients on $n$, the numerical relationship $\frac{C_{6i}^a}{2}=\frac{C_{6d}^c}{9}=C_{6d}^a$, and the large relative size of $C_{6i}^b$ ($\sim1000$ times larger than the others). These properties are explained by consideration of the relevant matrix elements, as described in the 
appendix.  Eq. \ref{eqn:interactionresults} presents the first calculation of the interaction between two Rydberg molecules, and  reveals the wide range of adjustable parameters and anisotropy present here, in marked contrast to Rydberg atoms. This stems from the molecular nature, since all of the coefficients and multipole moments depend on the internal structure and are not fixed properties of the atomic species. The van der Waals coefficients depend also on the field strength.

\begin{figure}[b]
{\normalsize 
\includegraphics[scale =0.145]{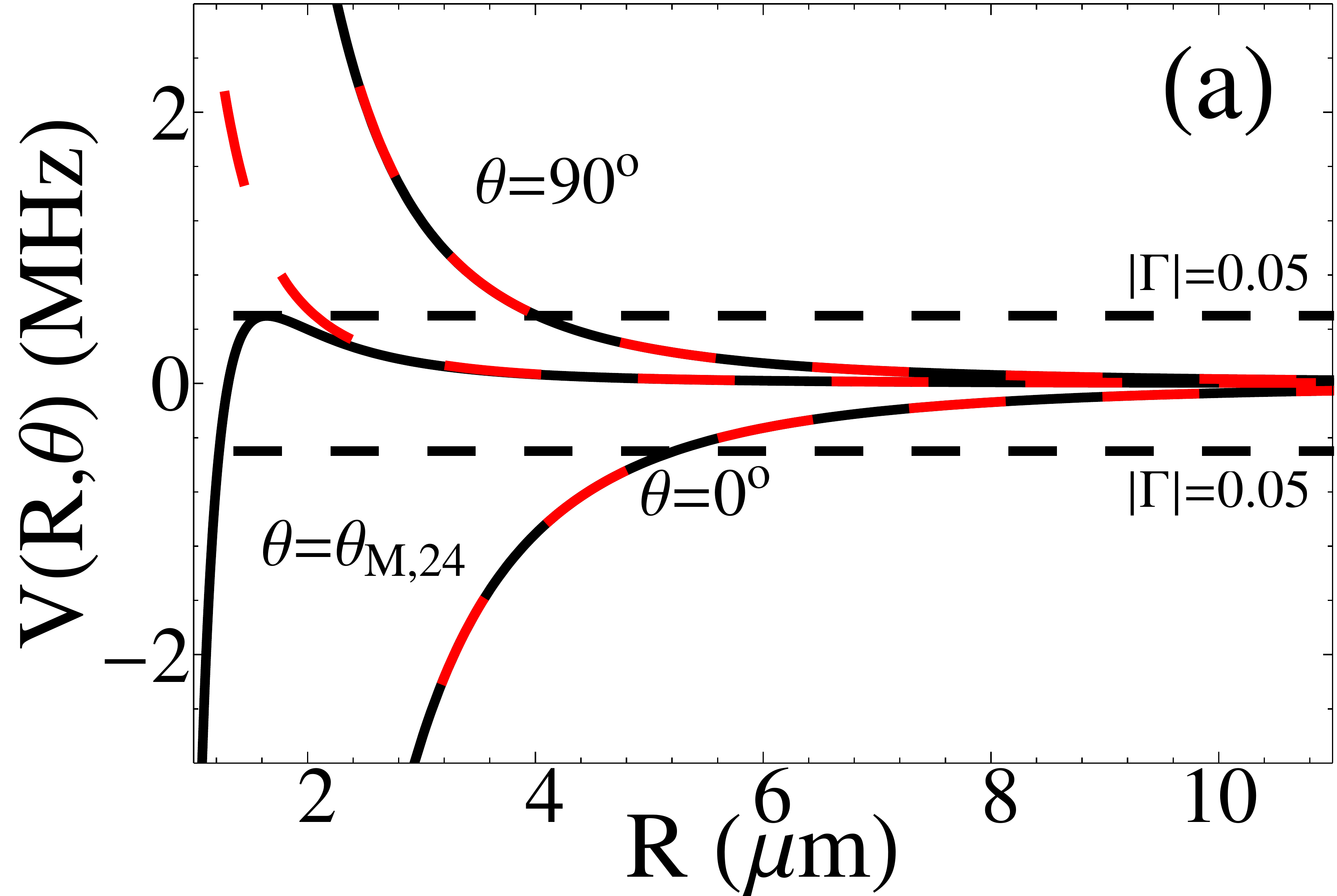}
{\vspace{-0pt}\includegraphics[scale =0.169]{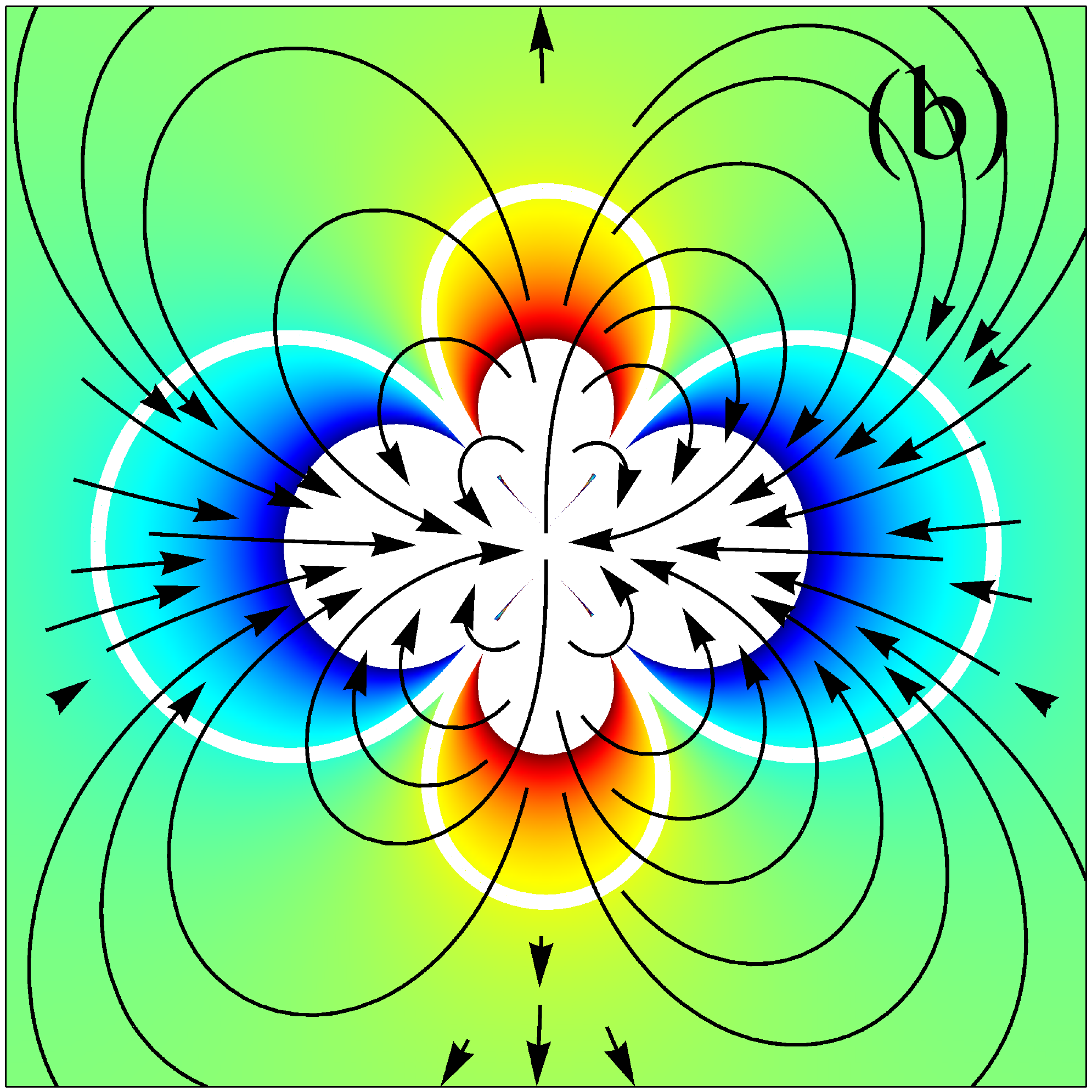}}
\vspace{-10pt}
}
\caption{(a) The interaction potential $V(R,\theta)$ for $n=24$ is plotted for three exemplary $\theta$ values. The pure dipole-dipole attraction is shown as in the dashed red curves, showing the importance of higher order terms near the magic angle. (b) The same potential is shown in cartesian coordinates, where blue(red) regions are attractive(repulsive). The inner white region (outer white contour) is the blockade radius satisfying $|V(R_B)|\ge \Gamma$, where $\Gamma = 0.5(0.1)$MHz. Lines of force are overlayed. The length of a side of the figure is $2\times 10^5$ a.u.    }
\label{fig:potentials}
\end{figure}

Fig. \ref{fig:potentials} displays the potential surface for $n=24$ in two different ways: in panel a three cuts of the potential surface at $\theta = 0^o,\theta_{M,24}$, and $90^o$ are plotted, while panel b displays a density plot of the potential. Lines of force are superimposed, and the inner(outer) white contours represent the blockade radius for $0.5(0.1)$MHz. At this scale this potential is largely determined by the dipole-dipole interaction, with  higher order effects playing a dominant role near $\theta_{M,n}$. This is seen in panel a, where the red dashed curves only include the DDI term, revealing that near the magic angle the higher order terms cause a repulsive barrier to form rather than the purely repulsive potential expected from the DDI alone. 

\section{Results and discussion}

\begin{figure}[tbp]
{\normalsize 
\includegraphics[scale =0.26]{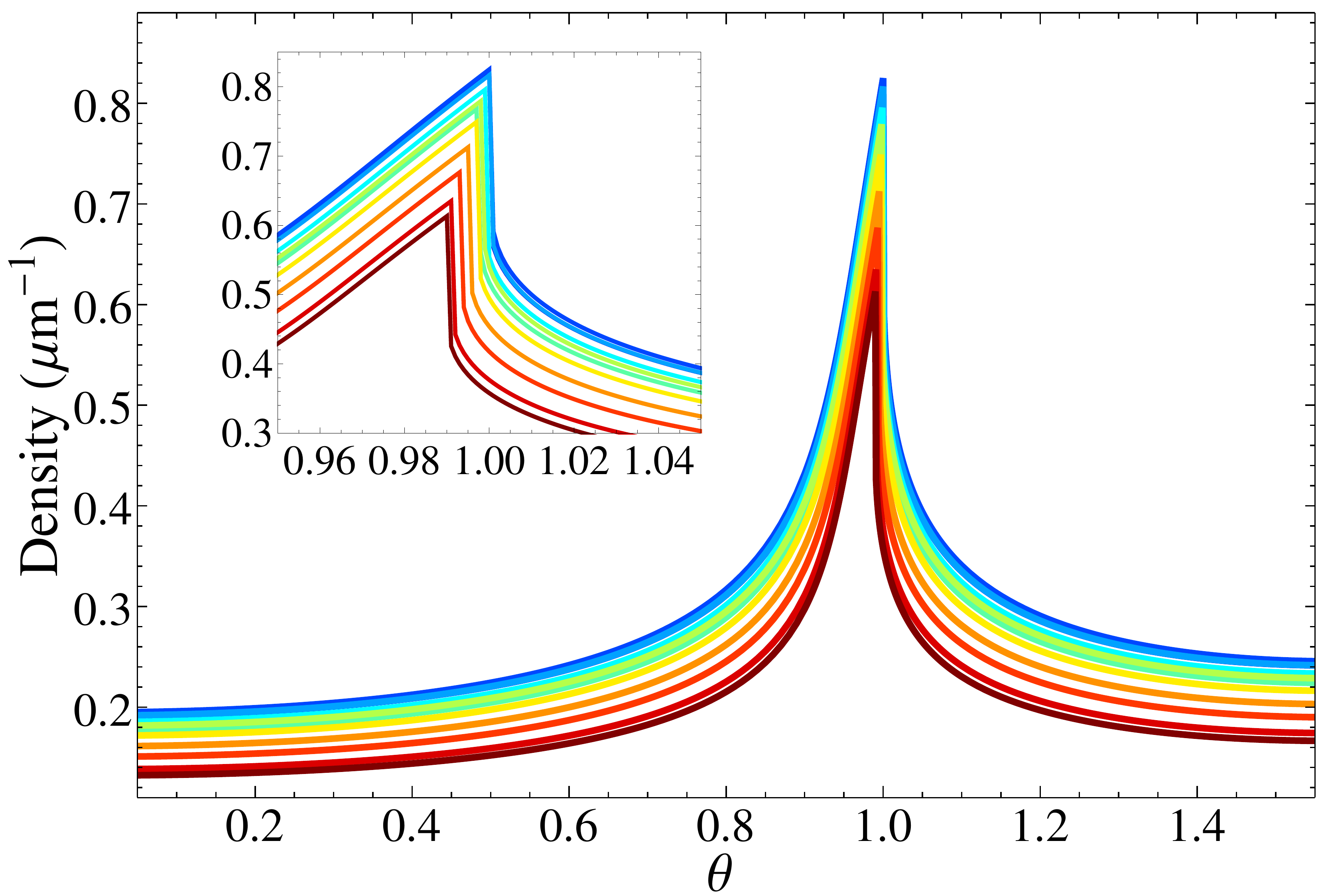}
\vspace{-10pt}
}
\caption{Simulated density as a function of the angle between the field and trap axes. As $n$ increases from 20-40 the color changes from blue to red.  The inset shows the density discontinuity for $\theta>\theta_{M_n}$.     }
\label{fig:ionyield}
\end{figure}

Fig. \ref{fig:ionyield} displays the key result of this article: the predicted density of butterfly molecules for a laser bandwidth $\Gamma = 0.5$MHz for several $n$ values. The $n$-dependent magic angle $\theta_{M,n}$ ranges from $56.7^\text{o}-57.3^\text{o}$ (0.99-1.0 radians) \footnote{For $n=(20,22,...,38,40)$ the values of $\Theta_{M,n}$ were $(57.3,57.3,57.3,57.2,57.1,57.2,57.1,57.0,56.9,56.8,56.7)$ degrees.} and differs slightly from the DDI magic angle $\theta_M=54.7^\text{o}$ due to the higher-order $n$-dependent terms in $V(R,\theta)$.  Decreasing with $n$ due to the stronger interactions at higher $n$, the density is $\sim4$ times greater at $\theta_{M,n}$ than at $\theta=0$ and corresponds to a maximum blockade radius of $\sim1\mu$m. The inset highlights the region near the $\theta_{M,n}$ to exaggerate the striking behavior here, in particular the discontinuity for $\theta$ barely greater than $\theta_{M,n}$. At this point the density drops discontinously. This stems from the interplay between the attractive van der Waals interactions and the DDI, as revealed in Fig. \ref{fig:potentials}a. When the latter interaction is repulsive, a potential barrier is created for $R$ greater than the inner crossing point where $V(R,\theta)=-\Gamma$. As the DDI increases relative to the van der Waals terms this repulsive barrier increases until it reaches $+\Gamma$, shifting the blockade radius suddenly.

The required atomic densities are large: a density of $2\times 10^{15}$cm$^{-3}$ gives two butterfly molecules per blockade volume. More favorable conditions are granted at higher Rydberg levels, where the interaction strengths are greater and the formation probability is enhanced due to the larger internuclear distance. A narrower bandwidth laser extends the effective range of the intermolecular interactions. Finally, an alternative setup could use a Q1D optical lattice in a doubly-occupied Mott insulator state~\cite{NiederpruemMott}. Other experimental schemes involving different classes of Rydberg molecules could also be studied: for example, the excitation of long-range ``trilobite" molecules through two-photon excitation of Cs$_2$ \cite{TallantCs,BoothTrilobite}. Bound by the s-wave scattering potential of Eq. \ref{fermieqn}, these molecules have larger bond lengths and dipole moments, allowing for more favorable density conditions. However, their lifetimes are shorter, and a two-photon process is required to access the $(n-4)S$ character of the trilobite admixture. Another scheme has been proposed in Ca, where trilobite-like molecules can be excited via two-photon excitation of the $nD$ state due to level shifts  from doubly excited states \cite{Eiles2015}. Even low-$l$ molecules exhibit weakly polar behavior from their small admixture of trilobite-sized dipole moments \cite{KurzPfau,PfauSci}. 

The dipolar length, which characterizes the length scale of the DDI, is  $a_{dd}=\frac{d^2m}{12\pi\epsilon_0\hbar^2}$ (in SI units), where $m$ is the mass of the particles. For the butterfly molecules, $a_{dd} \sim 10^7 - 10^8$a$_0$ (0.05 - 0.5 cm), whereas for the typical heteronuclear molecule KRb, $a_{dd}\sim1\mu m$. This overwhelming difference translates into huge dipolar interactions relative to ultracold collision interactions, characterized by the scattering length $a$ \footnote{The butterfly-butterfly scattering length is assumed to be the geometric size of the molecule, i.e., $a=4n^2$ a$_0$.}. This is characterized by the ratio $\varepsilon_{dd}=a_{dd}/a$, which is $\sim 10^4 -10^5$ for butterfly molecules and 20 for KRb. For magnetic atoms such as Cr or Dy $a_{dd}$ ranges from 16 to 130 $a_0$, giving $\varepsilon_{dd}\sim 1$ depending on the value of $a$. Following the work of Kalia and Vashishta~\cite{Kalia1981}, the present proposal leads to a crystal phase for butterfly molecules for the assumed $\sim\mu$K temperature. Indeed, the predicted one-dimensional array of molecules should resemble a linear crystal structure due to the fact that butterfly molecules are in the strongly interacting regime at ultracold temperatures. Moreover, the Tomonaga Luttinger liquid parameter~\cite{ReviewZoller}, given by $K=\rho a_{dd}$ in the strongly interacting regime~\cite{Citro2007,Citro2008}, is $\sim 10^3$ which clearly indicates the absence of density fluctuations and the prevalence of phase fluctuations.

Experimental realizations of polar Rydberg atoms are achieved by slow ramping of an electric field to induce a total adiabatic transition in an avoided crossing of the Stark manifold; this varies with different values of the electric field as well as for different atomic species. Moreover, these techniques generally involve a two-photon transition. The present approach is applicable to all atoms having an electron-neutral $P$-wave shape resonance and requires only a single photon transition. Furthermore, the dipole moment is an permanent property of the Rydberg molecule, stemming from the nature of the molecular bond itself, and  depends on its vibrational degrees of freedom; the applied static electric field is only used to align the existing dipole moments. 
\section{Conclusion}

We have presented an effective method to control the density of pendular butterfly molecules in a Q1D trap as a means of exploring the dipole-dipole interactions present in this system, and to explore the unique and exaggerated properties of this scheme contrasted with Rydberg atomic systems, polar molecules, or magnetic atoms. Future effort could explore the consequences of the inherently mixed nature of this system, consisting of dipolar impurities immersed in a sea of bosons, and thus could study polaron-polaron interactions in ultracold gases. The tunability of the polaron interaction provides information about the role of the internal structure of the impurities in the dynamics of the quasiparticles, which would elucidate the validity of the Fr\"ohlich Hamiltonian \cite{Frohlich} in the weak interaction regime and the study of the role of many-body correlations in the strong interaction regime, going beyond the single polaron physics very recently observed in ultracold gases \cite{Jorgensen, Hu}.  The newly developed theory of angulon and pendulon quasiparticles could be generalized to include impurity interactions, and could be realized in the present system \cite{Lemeshko1,Lemeshko2}. The generalization to two or three dimensional gases may help study the quantum phase transition between the superfluid and supersolid phases, or between supersolid and crystal phases \cite{ReviewZoller}. Future dynamical studies depend sensitively on the relationship between the molecular lifetimes and the relevant timescales for many-body physics. The already brief molecular lifetimes of several tens of microseconds may be reduced by interactions. However, the decay mechanisms are also interesting and largely unstudied. These decay mechanisms, such as Penning ionization, can be investigated using the anistropic interactions, since these can either be attractive, leading to collapse, or repulsive due to a potentia barrier, stabilizing the condensate.

\begin{acknowledgments}
This work is supported in part by the National Science 
Foundation under Grant No. NSF PHY-1607180 and Grant No. NSF PHY11-25915. M.T. Eiles and C.H. Greene are happy to acknowledge helpful discussions with G. Groenenboom.
\end{acknowledgments}

\appendix
\section{Calculation of the interaction potential using harmonic oscillator states}
To calculate the potential surface $V(R,\theta)$, the pendular states of a single molecule in the presence of an electric field are calculated, and then these states are used to calculate the interaction potential perturbatively. An accurate method for calculating pendular state eigenfunctions expands the rigid rotor Hamiltonian, $
H_\text{mol}
$, into the rotational basis of spherical harmonics $Y_{NM_{N}}(\theta,\phi)$, where $M_N$ is a good quantum number since the quantization axis is set parallel to the electric field. The pendular states, $\Psi_{\tilde NM_N}(\theta,\phi)=\sum_NC_{\tilde N,N}^{M_N}Y_{NM_N}(\theta,\phi)$, are characterized by their librational state, $\tilde N$. It proves convenient to define $W = E/B_e$ and $\omega = dF/B_e,\omega\gg 1$. Diagonalization of this matrix using values up to $N = 25-30$ gives a converged spectrum for the first $\sim 10$ excited states. The intermolecular interaction takes the form 
\cite{vanderAvoird1980}
\begin{align}
\label{eqn:interaction}
\hat V&=\sum_{L_A,L_B}q(L_A)q(L_B)\frac{f_{L_A,L_B}^{n}}{R^{L+1}}\sum_{m_A,m_B,m}\begin{pmatrix}L_A & L_B &L\\m_A & m_B & m\end{pmatrix}\nonumber\\\times &D_{m_A0}^{L_A}(\theta_A,\phi_A)^*D_{m_B0}^{L_B}(\theta_B,\phi_B)^*D_{m0}^{L}(\theta,0)^*,
\end{align}
where $L = L_A+L_B$ and $D_{m_X0}^{L_X}(\theta_X,\phi_X)^*$ is a Wigner D-Matrix rotating the multipole operator between the lab and molecule frame. $n$ is the principal quantum number and $q(L)$ is the $L$th reduced multipole moment. $D_{m0}^L(\theta,0)$ describes the geometry of the trap axis relative to the electric field axis, and 
\begin{align}
f_{L_A,L_B}^{n}&= (-1)^{L_A}n^{2L}\left[\frac{(2L+1)!}{(2L_A)!(2L_B)!}\right]^{1/2}.
 \end{align} 
 In first-order perturbation theory we include terms in Eq. \ref{eqn:interaction} up to $L = 4$, which includes quadrupole-quadrupole and dipole-octupole terms. To second order, we keep only the dipole-dipole ($L = 2$) term, since this contributes to order $1/R^6$. We also neglect retardation effects, falling off as $1/R^7$.  The dipole-dipole term is thus proportional to $D=n^8d^4/R^6$.  The dispersion and induction terms are calculated separately, where the summation in the induction term is just over the quantum numbers $\tilde N_B'$ and $M_B'$ of one molecule, while the dispersion term is summed over the virtual states of both molecules. 
\begin{align}
V_{ind}(R,\theta) &= D\sum\frac{|\langle\tilde N_AM_A\tilde N_BM_B|\hat V|\tilde N_AM_A\tilde N_B'M_B'\rangle|^2}{E_0 - E_{N_B',M_B'}}\\
V_{dis}(R,\theta) &= D\sum\frac{|\langle\tilde N_AM_A\tilde N_BM_B|\hat V|\tilde N_A'M_A'\tilde N_B'M_B'\rangle|^2}{2E_0 - E_{N_B',M_B'}- E_{N_A',M_A'}}.
\end{align}
 $E_0$ is the unperturbed energy of a single molecular state.  
 
 Greater insight into the character of these pendular states is given by considering the limit $\omega\to\infty$, since to lowest order in $1/\omega$ the Schr\"{o}dinger equation can be written as a two-dimensional harmonic oscillator. Using the explicit form for $\hat N^2$ in spherical coordinates, the Schr\"{o}dinger equation is
\begin{equation}
\left(\frac{\partial^2}{\partial\theta^2} +\cot\theta\frac{\partial}{\partial\theta} + \frac{1}{\sin^2\theta}\frac{\partial^2}{\partial\phi^2}+\omega\cos\theta + W\right)\Psi(\theta,\phi) = 0.\nonumber
\end{equation}
This equation maps onto the 2D Harmonic Oscillator by setting $\xi = 2\alpha\tan(\theta/2)$, where $\alpha = \sqrt{\omega/2}$. A separable solution in $\xi$ and $\phi$ is then obtained, where $\Psi(\xi,\phi) = U(\xi)\frac{1}{\sqrt{2\pi}}e^{im\phi}$, $m=|M|$. $U(\xi)$ is then given by:
\begin{align}
0&=  \left(1 + \frac{\xi^2}{4\alpha}\right)^2\left[\frac{d^2}{d\xi^2} + \frac{1}{\xi}\frac{d}{d\xi} - \frac{m^2}{\xi^2}\right] U(\xi)\\&+\frac{WU(\xi)}{\alpha}+ \frac{\omega}{\alpha}\frac{4 - \xi^2/\alpha}{4 + \xi^2/\alpha}U(\xi).\nonumber
\end{align}
Since in the pendular regime $\alpha\gg 1$, we discard all terms of order $1/\alpha$ to obtain the standard harmonic oscillator Schroedinger equation
\begin{equation}
\left[\left(\frac{d^2}{d\xi^2} + \frac{1}{\xi}\frac{d}{d\xi} - \frac{m^2}{\xi^2}\right) + \beta - \xi^2\right]U(\xi) = 0,
\end{equation}
where $\alpha\cdot\beta = W + \omega$. The energies of the pendular states are then given by
\begin{equation}
E = B_e(\sqrt{2\omega}(2\tilde N + |M| + 1)-\omega),
\end{equation}
 and the pendular states are
\begin{equation}
\Psi_{\tilde N,M}(\xi)=(-1)^M\sqrt{\frac{2\tilde N!}{\Gamma(\tilde N+M+1)}}e^{-\frac{\xi^2}{2}}\xi^ML_{\tilde N}^M(\xi^2),
\end{equation}
where $L_{N}^M(x)$ is a Laguerre polynomial. The accuracy of the large $\omega$ approximation is demonstrated in Fig. \ref{fig:pendular}. Using eq. \ref{eqn:interaction}, the matrix element connecting different two-molecule states is
 \begin{align}
 &\langle \tilde N_AM_A\tilde N_BM_B|\hat V|\tilde N_A' M_A'\tilde N_B'M_B'\rangle\\&= \nonumber\sum_{L_A,L_B}\frac{q(L_A)q(L_B)(4\pi)^{3/2}f_{L_A,L_B}^{n}}{R^{L+1}\sqrt{(2L+1)(2L_A+1)(2L_B+1)}}\\&\times\sum_{m_A,m_B,m}\begin{pmatrix}L_A & L_B &L\\m_A & m_B & m\end{pmatrix}Y_{L,m}(\theta,0)\nonumber\\&\times K_{\tilde N_AM_A,L_AM_A}^{\tilde N_A'M_A'}K_{\tilde N_BM_B,L_BM_B}^{\tilde N_B'M_B'},\nonumber
 \end{align}
 using $K_{\tilde N_AM_A,L_AM_A}^{\tilde N_A'M_A'}=\langle \tilde N_AM_A |Y_{L_A,m_A}(\xi,\phi)|\tilde N_A'M_A'\rangle$. 
 The first-order shift for the ground state simplifies to
\begin{align}
 &\langle 00,00|\hat V|00,00\rangle= -2\frac{d^2n^4}{R^3}P_2(x)\left(\frac{4\pi}{3}\left[K_{00,10}^{00}\right]^2\right)\\&-\frac{8n^8}{R^5}P_4(x)\left(\frac{4\pi}{\sqrt{21}}K_{00,10}^{00}K_{00,30}^{00}do-\frac{3\pi}{5}\left[K_{00,20}^{00}\right]^2\right).\nonumber
 \label{eqn:ground}
 \end{align}
 These three matrix elements can be found analytically in the harmonic oscillator approximation, and have relatively simple asymptotic forms for $\omega \gg 1$:
 \begin{align}
 \left(\frac{4\pi}{3}\left[K_{00,10}^{00}\right]^2\right)&\to 1 + \frac{3}{2\omega}  - \sqrt{\frac{2}{\omega}}\\
  \frac{3\pi}{5}\left[K_{00,20}^{00}\right]^2 &\to \frac{3}{8\omega}(21 - 6\sqrt{2\omega} + 2\omega)\\
    \frac{4\pi}{\sqrt{21}}K_{00,10}^{00}K_{00,30}^{00}&\to1 + \frac{14}{\omega}-\frac{7}{\sqrt{2\omega}}.
 \end{align}
 These saturate for large $\omega$, which is why the coefficients $C_3$, $C_{5a}$, and $C_{5b}$ can be taken as constant over the range of values studied here. For the second order dispersion terms, at the level of truncation discussed above, are
    \begin{align}
 &   V_d(R,\theta) = \frac{128\pi^3d^4n^8}{3R^6}\sum_{\tilde N_A,\tilde N_B}\sum_{|M_A|\le1,|M_B|\le 1}\\&\times \frac{\left[\begin{pmatrix}1 & 1 & 2\\-M_A & -M_B & M_A + M_B\end{pmatrix}K_{00,1-M_A}^{\tilde N_AM_A}K_{00,1-M_B}^{\tilde N_B M_B}\right]^2}{2E_0 - E_{\tilde N_A,M_A} - E_{\tilde N_B,M_B}},\nonumber
    \end{align}
    where $\tilde N_A = M_A =0$ and $\tilde N_B = M_B = 0$ terms are excluded from the sum. Before evaluating this expression to arbitrarily large $\tilde N_A$, $\tilde N_B$, we present the asymptotic forms of the relevant matrix elements ($ a= \sqrt{\frac{3}{8\pi}}$):
    \begin{align}
     K_{00,10}^{00} &\to a\left(\sqrt{2}-\frac{1}{\sqrt{\omega}}\right)\\
    K_{00,10}^{10}&\to a\left(-\frac{\sqrt{2}}{\omega} + \frac{1}{\sqrt{\omega}}\right)\\
    K_{00,1-1}^{01}&\to a\left(\frac{1}{(2\omega^3)^{1/4}}-\left(\frac{2}{\omega}\right)^{1/4}\right)\\
    K_{00,1-1}^{11}&\to a\left(\frac{1}{(2\omega)^{3/4}} - \frac{3}{(2\omega)^{5/4}}\right)\\
    K_{00,10}^{20}&\to a\left(\frac{1}{\sqrt{2}\omega} - \frac{9}{4\omega^{3/2}}\right)\\
    K_{00,1-1}^{21} &\to a\frac{\sqrt{3}}{2}\left(\frac{1}{2^{3/4}\omega^{5/4}} - \frac{3}{2^{1/4}\omega^{7/4}}\right).
     \end{align}
     These asymptotic forms suggest that the sum should converge rapidly, as all terms with final state in the $\tilde N = 2$ level are suppressed by additional factors of order $1/\sqrt{\omega}$. This is confirmed numerically; to better than 1\% accuracy the sum can be truncated to include only $\tilde N_A = \tilde N_B = 1$. After collecting into terms with the same angular dependence, this gives the first-order potential of Eq. \ref{eqn:interactionresults}, and gives explicit formulas for the coefficients:
    \begin{align}
    C_{6d}^a = \frac{8\pi^2}{9}&\Bigg(\frac{\left[K_{00,1-1}^{01}\right]^4}{2E_0 - 2E_{0,1}} + \frac{2\left[K_{00,10}^{10}\right]^4}{2E_0 - 2E_{1,0}} \\&\nonumber+ \frac{2\left[K_{00,1-1}^{01}\right]^2\left[K_{00,1-1}^{11}\right]^2}{2E_0 - E_{0,1} - E_{1,1}}+\frac{\left[K_{00,1-1}^{11}\right]^4}{2E_0 - 2E_{1,1}}\Bigg)
    \end{align}
    \begin{align}
    C_{6d}^b&= 32\pi^2\Bigg(\frac{\left[K_{00,10}^{10}\right]^2\left[K_{00,1-1}^{01}\right]^2}{2E_0 - E_{0,1} - E_{1,0}}+\frac{\left[K_{00,10}^{10}\right]^2\left[K_{00,1-1}^{11}\right]^2}{2E_0 - E_{1,0} - E_{1,1}}\Bigg)
    \end{align}
    \begin{align}
    C_{6d}^c= 8\pi^2& \Bigg(\frac{\left[K_{00,1-1}^{01}\right]^4}{2E_0 - 2E_{0,1}} + \frac{2\left[K_{00,1-1}^{01}\right]^2\left[K_{00,1-1}^{11}\right]^2}{2E_0 - E_{0,1} - E_{1,1}}\nonumber\\& + \frac{\left[K_{00,1-1}^{11}\right]^4}{2E_0 - 2E_{1,1}}\Bigg).
    \end{align}
    In this large $\omega$ regime, the only term making $C_{6d}^c=9C_{6d}^a$ not an exact relationship is $2\left[K_{00,10}^{10}\right]^4/(2E_0 - 2E_{1,0})$ in $C_{6d}^a$. This term declines rapidly with increasing $\omega$: rigorously computing the ratio $C_{6d}^c/C_{6d}^a$  for $\omega\gg 1$ limit shows the explicit $\omega$ dependence: 
    \begin{equation}
C_{6d}^c/C_{6d}^a\to9+9\sqrt{2/\omega^3}-9/(2\omega)
\end{equation}
We now calculate the induction term,
 \begin{align}
    V_i(R,\theta) &= \frac{128\pi^3d^4n^8}{3R^6}\sum_{\tilde N_B}\sum_{M_B|\le 1}\\&\times\nonumber\frac{\left[\begin{pmatrix}1 & 1 & 2\\0& -M_B & M_B\end{pmatrix}K_{00,10}^{00}K_{00,1-M_B}^{\tilde N_B M_B}\right]^2}{E_0  - E_{\tilde N_B,M_B}},
    \end{align}
    including only terms with $\tilde N_B \le 1$ for the same reasons given above. Once again, after collecting terms with the same angle-dependence Eqn. \ref{eqn:interactionresults} is found and the coefficients are defined 
    \begin{align}
    C_{6i}^a &= \frac{16\pi^2}{9}\frac{\left[K_{00,10}^{00}\right]^2\left[K_{00,10}^{10}\right]^2}{E_0 - E_{1,0}}\\
    C_{6i}^b &= 16\pi^2\left[K_{00,10}^{00}\right]^2\left(\frac{\left[K_{00,1-1}^{01}\right]^2}{E_0 - E_{0,1}} + \frac{\left[K_{00,1-1}^{11}\right]^2}{E_0 - E_{1,1}}\right).
    \end{align}
    Comparing $C_{6i}^a$ with $C_{6d}^a$, the ratio 16/8 = 2 is indicative of the relationship $C_{6i}^a = 2C_{6d}^a$, although it is not transparent how the different terms within the parentheses are equivalent. Once again, evaluating this ratio in the large $\omega$ limit reveals:
\begin{equation}
C_{6i}^a/C_{6d}^a=2-2\sqrt{2/\omega}.
\end{equation}
Finally, computing the ratios $C_{6i}^b/C$, where $C$ is any of the other coefficients, in the large $\omega$ limit shows that all of these ratios increase, at worst, as $\sqrt{\omega}$, explaining the large size of this coefficient compared to the rest. 

\bibliography{Ferro}
 
\end{document}